\documentclass[10pt,superscriptaddress,preprintnumbers,amsmath,amssymb]{revtex4}
\usepackage{epsfig} 
\usepackage{graphicx}
\usepackage{color}
\usepackage{float}
\restylefloat{table}

\large

\usepackage{amsfonts}
\usepackage{ulem}

\newcommand{\be}{\begin{equation}}
\newcommand{\ee}{\end{equation}}
\newcommand{\bea}{\begin{eqnarray}}
\newcommand{\eea}{\end{eqnarray}}
\newcommand{\ba}{\begin{array}}
\newcommand{\ea}{\end{array}}

\newcommand{\bla}{\color{black}}

\begin{document} 

\title{The Unruh effect interpreted as a quantum noise channel}

\author{S. Omkar}
\email{omkar.shrm@gmail.com}
\affiliation{Poornaprajna   Institute   of   Scientific   Research, Sadashivnagar, Bengaluru, Karnataka -  560080,  India}

\author{Subhashish Banerjee}
\email{subhashish@iitj.ac.in}
\affiliation{Indian Institute of Technology Jodhpur, Jodhpur 342011, India}

\author{R. Srikanth}
\email{srik@poornaprajna.org}
\affiliation{Poornaprajna   Institute   of   Scientific   Research, Sadashivnagar, Bengaluru, Karnataka -  560080,  India}

\author{Ashutosh Kumar Alok}
\email{akalok@iitj.ac.in}
\affiliation{Indian Institute of Technology Jodhpur, Jodhpur 342011, India}

\date{\today} 
\preprint{}

\begin{abstract}
We make use of the tools of quantum information theory to shed light
on the Unruh effect. A modal qubit appears as if subjected to quantum noise that
degrades quantum information, as observed in the accelerated reference frame. The
Unruh effect experienced by a mode of a free Dirac field, as seen by a relativistically
accelerated observer, is treated as a noise channel, which we term the “Unruh
channel”. We characterize this channel by providing its operator-sum representation,
and study various facets of quantum correlations, such as, Bell inequality violations,
entanglement, teleportation and measurement-induced decoherence under the effect.
We compare and contrast this channel from conventional noise due to environmental
decoherence. We show that the Unruh effect produces an amplitude-damping-like
channel, associated with zero temperature, even though the Unruh effect is associated
with a non-zero temperature. Asymptotically, the Bloch sphere subjected to the
channel does not converge to a point, as would be expected by fluctuation-dissipation
arguments, but contracts by a finite factor. We construct for the Unruh effect the
inverse channel, a non-completely-positive map, that formally reverses the effect, and
offer some physical interpretation. 
\end{abstract}

\maketitle 

\newpage

\section{Introduction}
Decoherence is an unavoidable  phenomenon associated with open quantum
systems,  which   occurs  because   of  their  interaction   with  the
environment. This causes the decay of quantum correlations, a resource
essential   for   quantum    information   processing.    A   detailed
understanding of decoherence, and  more generally, quantum information
in a  relativistic setting would  be relevant both from  a fundamental
perspective  as   well  as   to  help  future   experiments  involving
relativistic observers.

The relativistic effect named after Unruh \cite{Dav-Unr,crispino} predicts
that the  Minkowski vacuum  as seen by  an observer  accelerating with
constant proper  acceleration $a$ will  appear as a warm  gas emitting
black-body radiation at the \textit{Unruh temperature}:
\begin{equation}
\tau = \frac{\hbar a}{2\pi k_B c},
\label{eq:UT}
\end{equation}
where $c$  is the speed of  light in vacuum, and  $k_B$ is Boltzmann's
constant. The Unruh effect is usually calculated by idealizing the acceleration
as lasting for infinite time. However, in recent times, there have been
some interesting investigations exploring the finite acceleration aspects of
this effect \cite{bruschi2012} as well as the role of interaction time
of detectors with the quantum field \cite{hu-fukuma}.

The  Unruh effect produces a decoherence-like  effect, earning it
the  moniker `Unruh  channel'.   It degrades  the quantum  information
shared  between  an  inertial  observer  (Alice)  and  an  accelerated
observer (Rob), as seen in the  latter's frame, in the case of bosonic
or  Dirac field  modes  \cite{AM03, AFM+2006,  Tian2012}.   It may  be
combined with other,  non-relativistic noise, such as the  bit flip or
amplitude  damping  channels  \cite{SM2011,WJ2010}.  Being  conjugate
degradable, the  Unruh channel allows  for its quantum capacity  to be
computed efficiently  with a single-use of  the channel \cite{BDH+10}.
A channel is degradable  (conjugate degradable) if the environment can
be  simulated from  the  output  (up to  a  complex conjugation).  The
capacity of the Unruh channel for various resources such as classical,
quantum, public  or private communication  \cite{BHP09}, and trade-off
relations  between  such  capacities,  has  been  extensively  studied
\cite{BHP12,BHT+10} as part of  Hadamard channels in the bosonic case
\cite{CBW11,WHG12},  and  as  part   of  Grassmann  channels  in  the
fermionic  case  \cite{BCJ+11}.   The  Hadamard (Grassmann)  class  is
induced  by  the two-mode  squeezing  operator  for operators  obeying
bosonic (fermionic) algebra.

Differences  between the  two  cases arise  because  of the  different
algebras  and also  because  multi-mode fermions  lack tensor  product
structure. The Unruh channel is known to render separable an initially
maximally entangled state in the limit of infinite acceleration in the
case  of bosonic entanglement,  but not  so in  the case  of fermionic
entanglement \cite{AFM+2006}.  However, the channel capacities of both
channels  for  quantum  communication  between the  inertial  and  the
accelerated observer  are found to  be qualitatively similar,  in that
both vanish asymptotically \cite{BCJ+11}.

 The studies on Unruh effect form a part of the endeavour 
to understand relativistic aspects of quantum information \cite{czachor1997,peres1,caban1,caban2,terno1,avelar,
Bruschi:2012uf,Friis:2012cx,Bruschi:2012rx,Bruschi:2012rx,Lee:2014kaa,Sabin:2015wqa}, see for example the review
\cite{Peres:2002wx}. There have been analysis of Bell-CHSH kind of inequality for
relativistic spin systems \cite{caban3} as well as for photons \cite{caban4} 
and it was found that the relativistic quantum correlations could be stronger than nonrelativistic 
ones for a variety of configurations \cite{caban2}. There have also been suggestions to practical applications
in quantum information processing \cite{Bruschi:2012uf,Bruschi:2012rx}.
More recently, attention has also been directed towards
elementary particle physics \cite{particle-phy}, inspired by the technical
advances in high energy physics experiments, in particular
the meson factories and  neutrino experiments. The present work could be considered as a 
small contribution to these ongoing efforts in the field of relativistic quantum information.

  Although the above investigations show that the Unruh channel can be
  studied formally  as a class of  quantum noise channel, still  it is
  worth noting that  the Unruh channel does  not describe conventional
  quantum noise (which  arises because of a  system`s interaction with
  its environment).  Instead, it arises because the vacua perceived by
  an  inertial and  accelerated observer  do  not belong  to the  same
  Hilbert   space,  which   essentially   comes   about  because   the
  representations             of             the             canonical
  commutation/anti-commutation relations (CCR/CAR) of the
  quantum fields  in the  two vacua  are unitarily  inequivalent.  The
  Unruh  channel  thus   lies  in  some  sense  in  the   eye  of  the
  non-intertial  beholder. Here  we  shall see  that the  relativistic
  noise  seems  to be  somewhat  different  from its  non-relativistic
  counterpart.  An important non-relativistic noise channel, one which
  we will make use  of in this work as well,  is the amplitude damping
  (AD)   channel  \cite{NC00,sb1,sb2}.    This   channel  models   the
  dissipative influence of a zero temperature vacuum bath and is a sub
  class  of  a more  general  squeezed  generalized amplitude  damping
  (SGAD) channel \cite{sb1},  which takes into account  the effects of
  finite temperature and  squeezing. We will see below  that the Unruh
  channel, studied here, behaves like an interrupted AD channel.

 In this  work, we  study various aspects of quantum correlations, such as,
Bell inequality violations, entanglement, teleportation and measurement induced decoherence
under the effect of the Unruh channel. By providing a geometric characterization of the
Unruh channel, we highlight the points  of  similarity  and differences   between   
Unruh   channels   and   their conventional counterparts.  Works related to Unruh 
effects concentrate on either the bosonic \cite{FuentesSchuller:2004xp} or fermionic \cite{AFM+2006}
 channels, depending upon whether one is dealing with a scalar or a Dirac field, respectively.
It is known that in the bosonic case  entanglement degrades completely
in the infinite acceleration limit while in the fermionic case,
entanglement is never completely destroyed.  This motivates us to study the effect of 
other aspects of quantum correlations on the fermionic Unruh channel.

Linear maps describing the dynamic evolution of density operators of a
quantum system initially entangled with another quantum system are not
necessarily completely  positive (CP) \cite{AS2005,JSS04},  i.e., they
can map  a positive operator to  one that is not  positive, unless the
map's domain  is restricted. It  is possible to represent  such non-CP
maps in the usual operator-sum or Kraus form, with some negative terms
included, so that they are described as the difference of two CP maps.
To this end,  we first derive the operator-sum  representation for one
of  two modes  of a  Dirac field  described by  relatively accelerated
systems,  for e.g.,  detectors, in  a Minkowski  space time.   We then
compare and contrast the Unruh channel with conventional noise arising
from  environmental  decoherence,  in  order  to  obtain  a  physical
interpretation of it. 

The  article is organized  as follows:  In section  \ref{sec:unruh} we
introduce  the notion  of the  Unruh effect  for a  two-mode fermionic
system  and also  set up  the motivation  for an  Unruh  channel.  The
degradation of  quantum correlations caused by the  effect are studied
and quantified  by various measures  in section \ref{sec:degradation}.
In section  \ref{sec:geometry}, we characterize the Unruh  effect as a
quantum noise  channel.  We  compare and contrast  the latter  from an
AD  channel, and  discuss the  distortion  of the
Bloch  sphere under  its  influence. The  composability  of the  Unruh
channel with, and its  local indistinguishability from, an AD channel,
are discussed in  Section \ref{sec:comp}.  In section \ref{sec:rocket}
we  discuss,  from a  quantum  information  perspective, the  physical
interpretation of  the Unruh channel,  constructing an inverse Unruh map,
which is non-CP (NCP). Finally, we conclude in Section \ref{sec:unconclu}.

\section{Unruh effect\label{sec:unruh}}
The  Unruh  effect is  usually  studied  by examining  the
  Minkowski  (flat)   spacetime  in   terms  of   Rindler  coordinates
  \cite{rindler66}.  The Rindler  transformation  basically effects  a
  change  to cylindrical  coordinates in  Euclidean space  and divides
  spacetime  into  two  causally  disconnected wedges,  such  that,  a
  uniformily accelerated  observer in one wedge  is causally separated
  from the other  wedge. The fields in question, scalar  or Dirac, are
  quantized and expressed in terms  of linear combinations of creation
  and  annihilation  operators, for  both  the  Minkowski and  Rindler
  spacetimes.  Quantization  leads to  the  concepts  of particles  in
  either  spacetime. The  annihilation  operator of  particles in  one
  space time, say for example Minkowski spacetime, can be expressed in
  terms of  creation and  annihilation operators  of particles  in the
  Rindler  spacetime.  Thus  we  see  that  the  Minkowski  vacuum  is
  different  from the  Rindler  vacuum and  they  have different  Fock
  spaces.  The  linear coefficients  of the above  transformations are
  the        so        called        Bogoliubov        transformations
  \cite{bogoliubov1947,rindler-fulling}, relating the two Fock spaces.
  This leads  to the essence  of Unruh  effect formalized by  the term
  thermalization theorem \cite{davies} which essentially says that the
  average  number of  particles in  a uniformly  accelerated frame  as
  observed  in the  Minkowski vacuum  is given  by a  Bose or  a Fermi
  distribution function depending  upon whether one is  dealing with a
  scalar or Dirac field, respectively. 

Consider two observers, Alice $(A)$  and Rob $(R)$ sharing a maximally
entangled  state of two  Dirac field  modes (and  thus  a  qubit rather  than  qudit fermionic  Unruh
channel), at  a point  in Minkowski
spacetime, of the form
\begin{equation}
|\psi\rangle_{A,R}=\frac{|00\rangle_{A,R}+|11\rangle_{A,R}}{\sqrt{2}},
\label{eqn:unruh}
\end{equation} 
where $|j\rangle$ denote Fock states.  Let Rob move away from
stationary  Alice with a  uniform proper  acceleration $a$.  The
effect  of constant  proper  acceleration is  described  by a  Rindler
spacetime, which manifests two causally disconnected regions I and II,
where region I  is accessible to Rob, and separated  from region II by
an event horizon.

Using the above formalism, it can be shown that from Rob's
  frame the Minkowski vacuum state is a two-mode squeezed state, while
  the excited  state appears as  a product state,  see \cite{AFM+2006}
  for details:
\begin{eqnarray}
|0\rangle_M&\equiv&\cos r|0\rangle_I|0\rangle_{II}+\sin r|1\rangle_I|1\rangle_{II},\nonumber\\
|1\rangle_M &\equiv& |1\rangle_I|0\rangle_{II},
\label{eq:rindler}
\end{eqnarray}
\bla where   $\omega$   is  a   Dirac   particle   frequency  while   $\cos
r=\frac{1}{\sqrt{e^{-\frac{2\pi\omega   c}{a}}+1}}$  is  one   of  the
Bogoliubov coefficients,  connecting the Minkowski  and Rindler vacua.
It follows that $\cos  r \in[\frac{1}{\sqrt{2}},1]$ as $a$ ranges from
$\infty$  to 0.   As a  peculiarity of  this effect,
  we observe that  the states  in the  left hand  side of
Eq.  (\ref{eq:rindler}) are  single-mode  states, while  those in  the
right hand side are not.

Under      representation      (\ref{eq:rindler}),      the      state
Eq.   (\ref{eqn:unruh})   becomes   
\bea
|\psi\rangle_{A,I,II}   =
\frac{1}{\sqrt{2}}\left(|0\rangle_{A}(\cos r|0\rangle_I|0\rangle_{II}     +
\sin r|1\rangle_I|1\rangle_{II})  +|1\rangle_{A}|1\rangle_I|1\rangle_{II}\right).
\eea 
Tracing out mode II, we obtain the    density     matrix:     
\bea    
\rho^\prime_{A,R}=
\frac{1}{2}\left[\cos^2r|00\rangle\langle
  00|+\cos r(|00\rangle\langle11|+|11\rangle\langle
  00|)+\sin^2r|01\rangle\langle 01|+|11\rangle\langle
  11|\right],
\label{un:matrix}
\eea 
where the subscript $I$ has been replaced with subscript $R$\bla.
The  `evolution'   of  Rob's  qubit   to  a  mixed  state   under  the
transformation  $\mathcal{E}_U:\rho_{R}  \rightarrow  \rho^\prime_{R}$
constitutes what we call the Unruh channel for a fermionic qubit.

Here it would be pertinent to point out another interesting aspect of these
studies, viz. the Unruh modes \cite{funtes2010,Dav-Unr,birrell-davies} which are purely positive frequency linear
combinations of the Minkowski modes. This facilitates understanding the transformation between Minkowski and Rindler
modes by exploiting the fact that the connection between the Unruh and Rindler modes isolate the consequences of
the differing definitions of positive frequencies in the Minkowski and Rindler spacetime.  

\section{Degradation of quantum information under Unruh channel\label{sec:degradation}}

The  nonclassicality of  quantum information  can be  characterized in
terms   of  nonlocality   (the  strongest   condition),  entanglement,
teleportation fidelity or weaker nonclassicality measures like quantum
discord  or   measurement  induced   disturbance.   As  seen   in  the
accelerated reference frame, the Unruh effect degrades the quantumness
of the  state $\rho^\prime_{A,R}$ according  to each such  measure, as
discussed in the following subsections.

\subsection{Bell inequalities}

The violation of Bell-type inequality indicates that a given bipartite
state  $\rho$ cannot be  modelled using  a deterministic  local hidden
variable  (DLHV)   theory.   Suppose  Alice  holds   a  particle  with
dichotomic  properties  $A_1,  A_2  \in  \{\pm1\}$, and  Rob  holds  a
correlated particle  with similar properties  $B_1, B_2$. If  a hidden
variable $\lambda$ specifies the values of these four properties, then
the  quantity  $A_1(B_1-B_2)+A_2(B_1+B_2)=\pm2$.  Averaging  over  any
measure  of  $\lambda$, we  obtain  the the  Clauser-Horne-Shimony-Holt
(CHSH) inequality \cite{CS1978}
\begin{eqnarray}
\left| \langle A_1B_1\rangle-\langle A_1B_2\rangle+\langle A_2B_1\rangle+\langle A_2B_2\rangle\right|\leq 2,
\label{eq:CHSH} 
\end{eqnarray} 
where  $\langle  \dots\rangle$   stands  for  the  expectation  value.
Quantum mechanics  is nonlocal in  the sense that there  are entangled
states that  violate the CHSH inequality \cite{CHS+69, ADR1982}. On
the other hand, the violation may not be observed for a nonlocal state
if  the  settings  are  not optimal.   The  Peres-Horodecki  criterion
\cite{Per1996,HHH1996} helps here by giving a necessary and sufficient
condition for a bipartite quantum  state to be nonlocal.  Consider the
geometric  representation   for  a  general   two-qubit  mixed  state
\be
\rho=\frac{1}{4}[I_4+(\hat{r}\cdot\vec{\sigma})\otimes
  I_2+I_2\otimes       (\hat{s}\cdot\vec{\sigma})      +\sum_{i,j=1}^3
  \gamma_{ij}(\sigma_i\otimes \sigma_j)],
\label{eq:2bloch}
\ee
where  $I_4$ and  $I_2$ are identity  in four and  two dimensions,
respectively.   Let $\mu_1,\mu_2$ be  two eigenvalues  of $\Gamma^\dag
\Gamma$,  where $\Gamma$  is the  matrix $\{\gamma_{ij}\}$.   The CHSH
inequality  is  violated for some pairs of settings
of Alice and Rob if and only if 
\begin{equation} 
B(\rho)>1,
\label{eq:bell}
\end{equation} 
where $B(\rho)=\max(\mu_i+\mu_j)$,  \cite{HHH1996/2}. 
For $\rho'_{A,R}$,
$\Gamma^\dag\Gamma=
\textrm{Diag}(\cos^2r,\cos^2r,\cos^4r)$
 and   hence  $B(\rho)=2\cos^2r$.   For  $a\longrightarrow\infty$,
$\cos^2r\longrightarrow\frac{1}{2}$  and hence $B(\rho)\longrightarrow
1$.   In  Fig.   \ref{fig:un-srik}, the  quantity  $\frac{B(\rho)}{2}$
(which indicates nonlocality if greater than $\frac{1}{2}$) is plotted
as a function  of Unruh acceleration.   There it  is seen that the
state perceived by Rob becomes local around $a = 4.6$.

\subsection{Concurrence}

A weaker  measure of quantum correlation is  entanglement, for example
as quantified by concurrence for the general mixed state $\rho$ of two
qubits \cite{Woo1998} is given by
\begin{eqnarray}
C=\max(\lambda_1-\lambda_2-\lambda_3-\lambda_4,0),
\end{eqnarray}
where $\lambda_i$ is the square root of the eigenvalues, in decreasing
order,     of    the     matrix     
$\sqrt{\rho}(\sigma_y\otimes
\sigma_y)\rho^\ast(\sigma_y\otimes \sigma_y)\sqrt{\rho}$,
where $\rho^\ast$ is the complex conjugate of $\rho$, here $\rho'_{A,R}$, in the eigenbasis
of  $\sigma_z\otimes\sigma_z$.  In  Fig. \ref{fig:un-srik},  the black
curve  presents $C$  as a  function of  Unruh acceleration.   As $C>0$
asymptotically in Fig. \ref{fig:un-srik},  the Unruh channel is seen
to be  not entanglement-breaking.  Since the quantum  capacity of
the  Unruh  channel,  $Q(\mathcal{E}_U)$,  is  also  known  to  vanish
asymptotically, it appears that  the behavior of $Q(\mathcal{E}_U)$ is
closer  to  nonlocality than  to  entanglement,  a  point that  merits
further investigation. 

\subsection{Teleportation}

From the perspective of application to quantum information processing,
the quantumness  of correlations in $\rho'_{A,R}$ may  be quantified by
the  fidelity   between  the  input  and  output   states  of  quantum
teleportation \cite{BBC+1993} using the EPR channel $\rho'_{A,R}$. Here
it  may  be  noted  that  $\rho'_{A,R}$  can  also  be  a  mixed  state
\cite{Pop1994}, and not necessarily  pure, as required in the original
teleportation protocol.

Suppose Alice and Rob share  an initial mixed state $\rho'_{A,R}$ to be
used  as channel  for teleportation,  and Alice  has an  unknown state
$|\psi\rangle$  of a third  particle $A^\prime$,  to be  teleported to
Rob. If  $\mathcal{B}_k$ denotes the  projectors to the Bell  basis in
the Hilbert  space $H_{A,A^\prime}$, then the probability  to obtain a
given    Bell   basis    outcome   $k$    is   $q_k    =   \textrm{Tr}
[(\mathcal{B}_k\otimes       I)(|\psi\rangle\langle\psi|       \otimes
  \rho'_{A,R})]$.   Based on  Alice's classical  commutation of  $k$ to
Rob,  he  applies a  suitable  local  unitary  $V_k$ to  obtain  state
$\rho_k$. If $|\psi\rangle$ is  chosen uniformly from the Bloch sphere
then the fidelity of the  input and output states of the teleportation
is given by
\begin{eqnarray}
F=\int_S d_H\psi\sum_k q_k\textrm{Tr}(\rho_k |\psi\rangle\langle\psi|),
\end{eqnarray}
where the integral  uses a uniform Haar measure.   Maximizing over all
quadruples     of    such     $V_k$    gives     \cite{HHH1996}    
\be
F_{\max}=\frac{1}{2}\left(1+\frac{1}{3}\left(\textrm{Tr}\sqrt{\Gamma^\dag\Gamma}\right)\right),
\ee 
where  $\mu_j$ are  eigenvalues of $\Gamma^\dag\Gamma$.
   The maximum fidelity attainable  in the above
teleportation protocol without the use of entanglement is known to be
$\frac{2}{3}$. Thus when $F_{\max}>\frac{2}{3}$, genuine teleportation
occurs,  and  the  state  $\rho'_{A,R}$  is  said  to  contain  quantum
correlations.

For  $\rho'_{A,R}$, $F_{\max}=\frac{1}{2}\left(1+\frac{1}{3}\left(2\cos
r+\cos^2r\right)\right)$.  In Fig. \ref{fig:un-srik}, the large-dashed
curve presents $F_{\max}$ as a function of Unruh acceleration.

\subsection{Measurement induced disturbance}

\begin{figure}
\begin{center}
\includegraphics[width=0.8\linewidth]{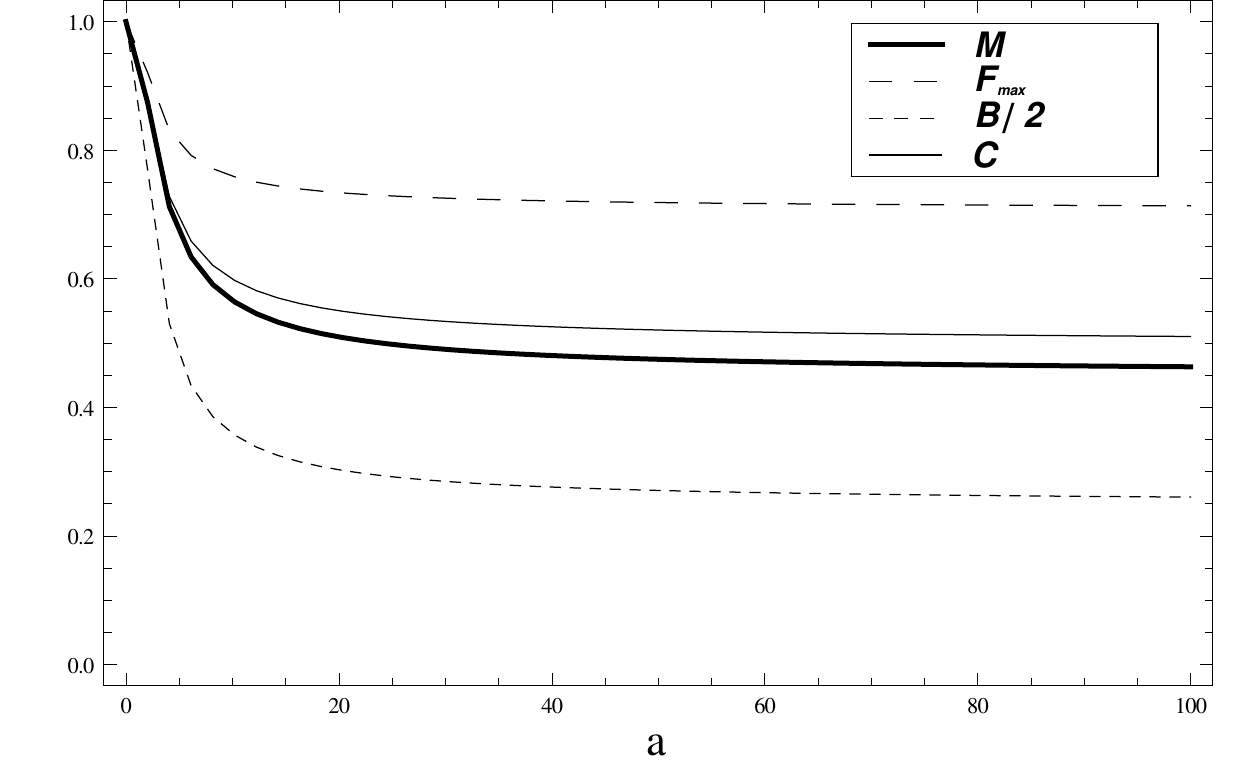}
\caption{Degradation    of   QMID   ($M$),    teleportation   fidelity
  ($F_{\max}$),  Bell  quantity ($B/2$)  and  concurrence  ($C$) as  a
  function  of Unruh  acceleration ($a$),  for $\omega=0.1$  (in units
  where  $\hbar\equiv c\equiv1$).  The  system becomes  local ($B/2  <
  1/2$) at  $a\approx4.6$, but stays nonclassical with  respect to the
  other parameters ($C > 0,F>\frac{2}{3}, M>0)$).}
\label{fig:un-srik}
\end{center}
\end{figure}

Measurement  induced  disturbance (QMID) quantifies  the  quantumness of  the
correlation between the quantum  bipartite states shared amongst Alice
and Rob. For the given  $\rho'_{A,R}$, if $\rho'_A$ and $\rho'_R$ are the
reduced density matrices, then  the mutual information that quantifies
the     correlation     between    Alice     and     Rob    is    
 \be
I=S(\rho'_A)+S(\rho'_R)-S(\rho'_{A,R}),  
\ee 
where $S(.)$  represents von
Neumann  entropy.  If   $\rho'_A=\sum_i  \lambda_A^i  \Pi_A^i$  and 
$\rho'_R=\sum_j \lambda_R^j \Pi_R^j$ denotes the spectral decomposition
of $\rho'_A$  and $\rho'_R$,  respectively, then the  state $\rho'_{A,R}$
after   measuring   in    joint   basis   $\{\Pi_A,\Pi_R\}$   is   
\begin{equation}
\Pi(\rho^\prime_{A,R})=\sum_{i,j}(\Pi_A^i\otimes\Pi_R^j)\rho^\prime_{A,R}
(\Pi_A^i\otimes\Pi_R^j).  
\end{equation}  
Since  there   exists  a   local  measurement  strategy   that  leaves
$\rho'_{A,R}$ unchanged,  $\rho'_{A,R}$ can be  considered as classical.
Then, $I(\Pi(\rho'_{A,R}))$ quantifies  the classical correlation. Thus
the difference between $I(\rho'_{A,R})$ and $I(\Pi(\rho'_{A,R}))$ should
quantify the  quantumness of correlation between Alice  and Rob.  This
difference           known            as           QMID,          
 \be
M(\rho^\prime_{A,R})=I(\rho^\prime_{A,R})-I(\Pi(\rho^\prime_{A,R}))
\ee  
is  a measure  of
quantumness of  the correlation. In  Fig. \ref{fig:un-srik}, we
find  that $M>0$ throughout  the range  considered, implying  that the
system remains nonclassical, as expected. 

\section{Geometric characterization of the Unruh channel\label{sec:geometry}}  

Since  in this  work we  are applying  quantum information  processing
ideas  to the  Unruh  channel,  it would  be  interesting  to have  an
understanding of  this channel using information  theoretic tools.  To
this end we invoke the Choi-Jamiolkowski isomorphism \cite{choi,jamio}
which is a two-way mapping between a quantum state and quantum channel
and is  an expression of channel-state  duality.  Information, quantum
as  well as  classical, is  transmitted via  a quantum  channel, which
basically is  a completely positive  map between spaces  of operators.
One way to obtain the state from the channel is to apply the latter on
one half  of a  maximal two-qubit or  two-qudit entangled  state.  The
resulting state is called the Choi matrix.  In the converse direction,
a unique  qudit channel can be  associated with each such  Choi matrix
via the idea of gate  teleportation \cite{vidal}.  The Kraus operators
for  the channel  can be  obtained  by diagonalizing  the Choi  matrix
\cite{debbie},   as  well   as  from   the  dynamical   map  for   the
transformation  \cite{rodrig,usha}.   Once  the  Kraus  operators  are
obtained, the channel gets completely characterized. We now proceed to
apply this to the Unruh channel.

Consider the  maximally entangled two-mode  state in which  the second
mode  is Unruh  accelerated. The  resulting state,  which is  the state
given in Eq. (\ref{un:matrix})  is 
\begin{equation} 
\rho_U=\frac{1}{2}\left(
 \begin{array}{cclr} \cos^2r&0&0&\cos r\\ 0&\sin^2r&0&0\\ 0&0&0&0\\
\cos r&0&0&1
 \end{array}\right).
\label{eq:uchoi}\ee  
Without   the   factor   $1/2$,    $\rho_U$   is   the   Choi   matrix
$\sum_{j,k}|j\rangle\langle  k|\otimes  \mathcal{E}_U(|j\rangle\langle
k|)$  corresponding  to  Unruh  channel  $\mathcal{E}_U$.   Spectrally
decomposing       this,        suppose       we       obtain       \be
\rho_U=\sum_{j=0}^3|\xi_j\rangle\langle\xi_j|, 
\end{equation}  
where $|\xi_j\rangle$ are the eigenvectors  normalized to the value of
the  eigenvalue.   Then  by Choi's  theorem  \cite{choi,debbie},  each
$|\xi_j\rangle$ yields a Kraus operator  obtained by folding the $d^2$
(here: 4)  entries of the  eigenvector in to $d\times  d$ ($2\times2$)
matrix, essentially  by taking each sequential  $d$-element segment of
$|\xi_j\rangle$, writing  it as a  column, and then  juxtaposing these
columns to form the matrix \cite{debbie}.

The two eigenvectors corresponding to the two non-vanishing eigenvalues
are found to be 
\bea
|\xi_0\rangle&=&(\cos r,0,0,1)\nonumber\\
|\xi_1\rangle&=&(0,\sin r,0,0).
\eea
From these,  we have  the  Kraus representation  for  $\mathcal{E}_U$ as  
\be
K^U_1=\left(
\begin{array}{cclr}
\cos r&0\\
0&1
\end{array}\right);~~
K^U_2=\left(
\begin{array}{cclr}
0&0\\
\sin r&0
\end{array}\right),
\label{eq:relcha}
\end{equation}
whereby
\begin{equation}
\mathcal{E}_U(\rho) = \sum_{j=1,2} K^U_j \rho \left(K^U_j\right)^\dag,
\label{eq:unrucha}
\end{equation}
with the completeness condition
\begin{equation}
\sum_{j=1,2} \left(K^U_j\right)^\dag  K^U_j = \mathcal{I}.
\end{equation}
This  is  formally similar  to  the  operator  elements in  the  Kraus
representation of an  AD channel \cite{NC00}, which  models the effect
of  a  zero temperature  thermal  bath  \cite{NC00,sb1,sb2}.  This  is
surprising as the Unruh effect corresponds to a finite temperature and
would naively be expected to correspond to the \textit{generalized} AD
or SGAD  channels, which  are finite temperature  channels. This  is a
pointer  towards,  as we  shall  see  in Section  \ref{sec:rocket},  a
fundamental difference  between the  Unruh and  the AD  channel. 
  Here it is interesting to note that the relativistic channel arising
  from  the expanding  universe  described by  a 1+1  Robertson-Walker
  spacetime,  is  also  known  to   be  analogous  to  an  AD  channel
  \cite{MPW+14}.

For an initial pure qubit state 
$\rho=
|0\rangle\langle0|\cos^2\frac{\theta}{2}
+ |0\rangle\langle1|e^{i\phi}\cos\frac{\theta}{2}\sin\frac{\theta}{2} +
|1\rangle\langle0|e^{-i\phi}\cos\frac{\theta}{2}\sin\frac{\theta}{2}
+ |1\rangle\langle1|\sin^2\frac{\theta}{2},$
the action of the Unruh channel is
\begin{equation}
\mathcal{E}_U(\rho)= \left(\begin{array}{clclr}
\cos^2r\cos^2\frac{\theta}{2}& \cos r e^{i\phi}\cos\frac{\theta}{2}\sin\frac{\theta}{2}\\
\cos r e^{-i\phi}\cos\frac{\theta}{2}\sin\frac{\theta}{2}& \sin^2r\cos^2\frac{\theta}{2} +
\sin^2\frac{\theta}{2}
\end{array}\right)\,.
\label{eq:rhocha}
\end{equation}

\begin{figure}
\begin{center}
\includegraphics[width=.6\linewidth]{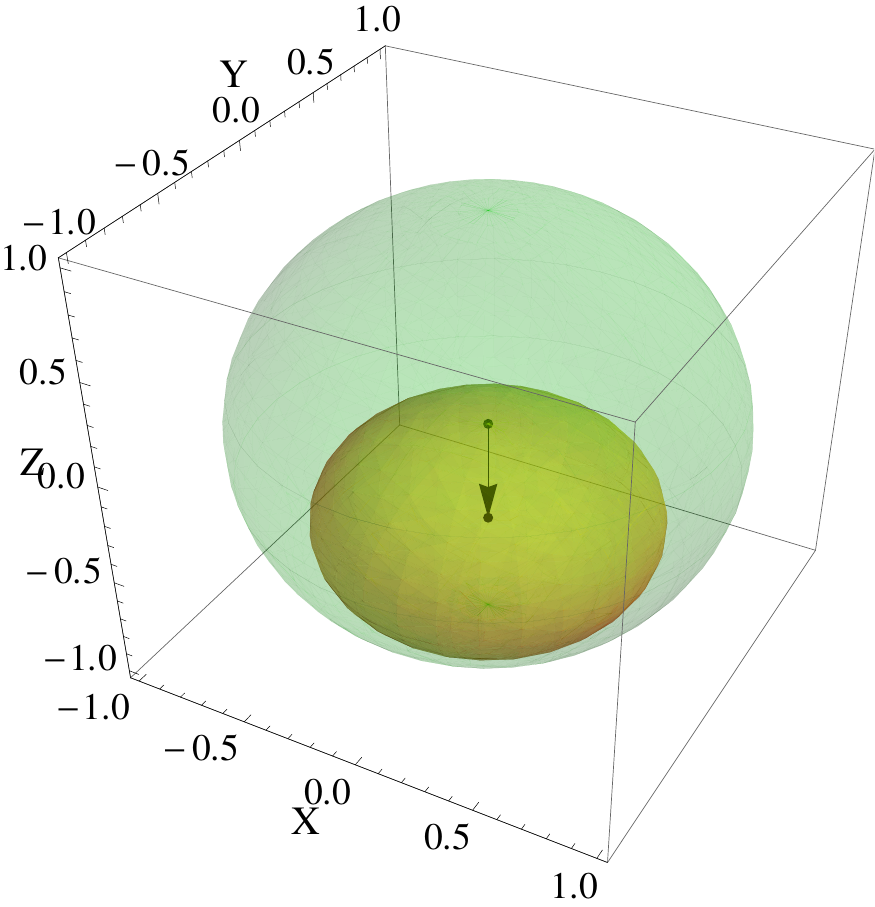}
\caption{Under $\mathcal{E}_U$, the Bloch sphere (the outer sphere) is
  shrunk   asymptotically   to  the   inner   solid  oblate   spheroid
  (eccentricity    $e=\frac{1}{\sqrt{2}}$)   by   a    volume   factor
  $\frac{1}{4}$,  centered at $(0,0,-\frac{1}{2})$  (surface described
  by $\hat{n}^\infty(\theta,\phi)$.}
\label{fig:accball}
\end{center}
\end{figure}

For infinite time and acceleration, setting $\cos^2r = \frac{1}{2}$ in
Eq. (\ref{eq:rhocha}), the asymptotic state is
\begin{equation}
\mathcal{E}_U(\rho)= \left(\begin{array}{clclr}
\frac{1}{2}\cos^2\frac{\theta}{2}&  \frac{1}{\sqrt{2}}e^{i\phi}\cos\frac{\theta}{2}\sin\frac{\theta}{2}\\
\vspace{.2cm}
\frac{1}{\sqrt{2}} e^{-i\phi}\cos\frac{\theta}{2}\sin\frac{\theta}{2}& \frac{1}{2}\cos^2\frac{\theta}{2}+\sin^2\frac{\theta}{2}
\end{array}\right),
\end{equation}
with   Bloch    vector       given   by   the    function:
\be
\hat{n}^\infty(\theta,\phi)\equiv
(\hat{x},\hat{y},\hat{z})= \left(\frac{\cos\phi\sin\theta}{\sqrt{2}},
\frac{\sin\phi\sin\theta}{\sqrt{2}},-\sin^2\frac{\theta}{2}\right),
\label{eq:bloch}
\ee 
which shows that  the Bloch sphere gets mapped to
the inscribed  solid object  shown in Fig.  \ref{fig:accball}, whose
south pole  osculates with that of  the Bloch sphere,  while the north
pole (corresponding  to initial $\theta=0$) is  located midway between
the  Bloch sphere  center  and south  pole: $\hat{n}^\infty(0,\phi)  =
(0,0,0)$ while  $\hat{n}^\infty(\pi,\phi) = (0,0,-1)$. This  is thus a
kind of an interrupted AD channel.  By virtue of linearity of the map,
it follows  that the  maximally mixed state  maps to the  Bloch vector
which is the average of the above two, being
\begin{equation}
\hat{n}^\infty(\mathcal{I}) = (0,0,-\frac{1}{2}).
\label{eq:offc}
\end{equation}
Thus the channel  is non-unital, with the new  Bloch representation of
the     initially     maximally      mixed     state     being    
 \be
\mathcal{E}_U^{\infty}(\mathcal{I}/2)=\left(\begin{array}{clclr}
  1/4&0\\ 0&3/4
\end{array}\right).
\end{equation} 
The geometry of the contracted,  noisy version of the Bloch sphere can
be inferred 
to be
\begin{equation}
R(\theta)=\left|\hat{n}^\infty(\theta,\phi) - \hat{n}^\infty(\mathcal{I})\right|
= \frac{\sqrt{3-\cos2\theta}}{2\sqrt{2}}.
\label{eq:acrad}
\end{equation}
From $\hat{n}^\infty(\theta,\phi)$, the  area of the  circular section  of the
oblate      spheroid      is      $A(\theta)=\pi(\hat{x}^2+\hat{y}^2)=
\frac{\pi}{2}\sin^2\theta$.  The  corresponding volume is  then, using
the    $z$-component   in    $\hat{n}^\infty(\theta,\phi)$,   $V=\int_0^\pi
A(\theta)\frac{\sin\theta}{2}  d\theta   =\frac{\pi}{3}$.   Since  the
volume of the Bloch sphere  is $V_0 \equiv \frac{4\pi}{3}$, it follows
that the volume contraction  factor of the
Bloch  sphere  under  the relativistic channel is $K\equiv\frac{1}{4}$.
The eccentricity of the oblate sphere is given by 
\begin{equation}
e = \sqrt{1 - \frac{a^2}{b^2}} = \frac{1}{\sqrt{2}},
\label{eq:ecc}
\end{equation}
where $a$  and $b$ are the  semi-minor and semi-major  axis, which are
seen   from the form of $R(\theta)$  to   be   $\frac{1}{\sqrt{2}}$
(corresponding    to    $\theta=\frac{\pi}{2}$)   and    $\frac{1}{2}$
(corresponding to $\theta=0,\pi$), respectively.

\section{Composability of AD and Unruh channels \label{sec:comp}}
 An noted in the  introduction, an important noisy channel
  is  the  AD  channel to  which the  Unruh  channel  is  similar.
  However, as seen  below, from a geometrical  characterization of the
  Unruh  channel, we  find  that  it behaves  like  an interrupted  AD
  channel.

 The similarity of the  Unruh channel $\mathcal{E}_U$ to the AD
channel  means  that, for sufficiently small damping, 
from  local  observations  on  the degradation  of
quantum information, one  cannot say whether one is  in an accelerated
frame and looking  at Minkowski vacuum, or in an inertial frame,
interacting dissipatively  with a Minkowski vacuum.  As  one aspect of
this equivalence, a genuine  AD channel may be composed with an Unruh channel,
resulting in an AD channel.  Consider  a state 
\be 
\rho=\left(\begin{array}{clclr} 
\alpha &  \beta \\ 
\beta^\ast&1-\alpha
\end{array}\right),
\ee
with real $\alpha$. The action of Unruh noise yields
\be
\rho_U=\mathcal{E}_U(\rho)=\left(\begin{array}{clclr}
\alpha\cos^2r&\beta\cos r\\
\beta^\ast\cos r& 1 -  \alpha\cos^2r 
\end{array}\right),
\ee
while that of an AD channel on state $\rho_U$ above is
\be
\mathcal{E}_{UV}(\rho) =
\left(\begin{array}{clclr}
\alpha\cos^2r(1-\gamma)&\beta\cos r\sqrt{1-\gamma}\\
\beta^\ast\cos r\sqrt{1-\gamma}& 1 + \alpha(-\cos^2r+\gamma \cos^2r) 
\end{array}\right)\,.
\ee
Setting $\gamma^{\prime\prime}=\gamma\cos^2r+\sin^2 r$,
the Kraus operators of noise $\mathcal{E}_{UV}$ are 
\be
\left(\begin{array}{clclr}
\sqrt{1-\gamma^{\prime\prime}}&0\\
0&1 
\end{array}\right),	~	~
\left(\begin{array}{clclr}
0&0\\
\sqrt{\gamma^{\prime\prime}}&0 
\end{array}\right),
\ee 
which has  the form of an AD channel.   The same observation holds
if the order of the AD  and Unruh channels are inverted.  This closure
under composition is a manifestation  of the semi-group property of AD
channels.

\section{Physical interpretation\label{sec:rocket}}

The origin of  the Unruh channel is, as noted,  quite different from a
conventional  noise channel.  We  discuss below  some aspects  of this
difference from a quantum information theoretic perspective.


In  spite of  the formal  similarity  of the  Unruh effect  to the  AD
channel, there are two differences from a conventional AD channel. One
is that the  damping parameter $\sin r$ in  Eq.  (\ref{eq:relcha}) can
never  go  to  1.   This   means  that  the  state  remains  logically
reversible,  and Rob can  in principle  reconstruct the  initial state
tomographically.   For  an AD  channel,  which  is  derived using  the
Born-Markov  and  rotating-wave  approximations,  one can  invoke  the
fluctuation-dissipation  theorem  to require  the  convergence of  the
Bloch sphere at initial time to  a single fixed point of the evolution
asymptotically.   The  finite contraction  factor  derived in  Section
\ref{sec:geometry} thwarts this behavior.

Another point is that the rank of $\rho_U$ in Eq. (\ref{eq:uchoi}) is
2, and  not 4. This  is reflected  in the fact  that there are  only 2
canonical Kraus  operators, given in Eq.   (\ref{eq:relcha}).  This is
somewhat  surprising because  we  ascribe a  finite Unruh  temperature
$\tau$  to the  bath observed  by the  accelerated observer,  which is
expected  to  correspond to  $\rho_U$  of  rank 4  \cite{OSB13}.   This
entails that  the Unruh  noise corresponds to  only a  single Lindblad
channel  corresponding  to  a   de-excitation  process.   Now  if  the
environment  were  a  conventional finite-temperature  bath,  then  we
should have also the Lindblad  excitation channel corresponding to the
qubit absorbing  a photon from this  bath. The lack of  the excitation
channel  here  suggests that,  from  the  physical perspective  of  an
inertial detector,  the Unruh background  interacts as a  vacuum, even
though Rob  views it as a  thermal Rindler state in  his own reference
frame. 

We can formally construct an  inverse Unruh channel $\Phi^{-1}_U$ such
that
\begin{equation}
\Phi^{-1}_U\left[\Phi_U(\rho)\right] = \rho.
\label{eq:solve}
\end{equation}
This cannot  be CP because, given  that the Unruh channel  proceeds by
setting up  an entanglement  between the  system and  its environment;
  $\Phi^{-1}$  would  be  a  map on  a  state  that  is  initially
entangled, making it in general an NCP map.  Solving (\ref{eq:solve}),
we  find   that  the  required   inverse  channel  is  given   by  the
\textit{difference}  of  two  CP   (but  non-trace  preserving)  maps,
determined   by  the   single  Kraus   operators,  respectively:   
\begin{equation}
K^\epsilon_1=\left(
\begin{array}{cclr}
1/\cos r&0\\
0&1
\end{array}\right)\,,~~
K^\epsilon_2=\left(
\begin{array}{cclr}
0&0\\
\tan r&0
\end{array}\right)\,.
\label{eq:relcha0}
\end{equation}
Thus, the inverse-Unruh channel
\begin{equation}
\Phi^{-1}_U(\rho)  = K^\epsilon_1\rho K^\epsilon_1  - K^\epsilon_2\rho
K^\epsilon_2,
\label{eq:unun}
\end{equation}
where  the   intervening  \textit{negative}  sign   is  the  tell-tale
indication of  $\Phi^{-1}_U$ being a  non-CP map, a  general Hermitian
map that can formally apply even when the initial correlations between
the system and the bath are non-classical \cite{SL09,JSS04}.

It  may  be verified  that  $\Phi^{-1}_U$  satisfies the  completeness
condition:
\begin{equation}
\left(K^\epsilon_1\right)^\dag K^\epsilon_1 - \left(K^\epsilon_2\right)^\dag K^\epsilon_2
= \mathcal{I}\,.
\label{eq:complete}
\end{equation}
That the inverse-Unruh channel $\Phi^{-1}_U$  is not CP can be seen by
its    action    on    the maximally    entangled    state    $\Psi$    :
\begin{equation}
(\Phi^{-1}_U\otimes        I)       \Psi        =       \frac{1}{2}
\left(\begin{array}{clclr}                         1/\cos^2r&0&0&1/\cos
  r\\ 0&-\tan^2r&0&0\\ 0&0&0&0\\ 1/\cos r&0&0&1
\end{array}\right)\,.
\end{equation}     
The       corresponding      eigenvalues       are      $\{0,0,(3+\cos
2r)/(4\cos^2r),-\tan^2r/2\}$.    Since   one    of   the   eigenvalues
$-\tan^2r/2$ is negative, $\Phi^{-1}_U$ is  an NCP map. Also note that
$\textrm{Tr}((\Phi^{-1}_U\otimes  I) \Psi)=1$,  a  consequence of  Eq.
(\ref{eq:complete}).  Thus  the $\Phi^{-1}_U$ is trace  preserving but
not completely positive.

What is  the physical  significance of  $\Phi^{-1}_U$? Conventionally,
NCP maps  are regarded  as unphysical,  since they  cannot be  given a
traditional  Kraus representation.  Can $\Phi^{-1}_U$  be regarded  as
describing  the transformation  corresponding to  ``turning off''  the
Unruh drive? That does not seem plausible, since the Unruh effect is
calculated idealizing the acceleration as lasting for infinite time.

\section{Discussion and conclusions \label{sec:unconclu}}

 In this  work, we  study various aspects of quantum correlations, such as,
Bell inequality violations, entanglement, teleportation and measurement induced decoherence
under the effect of the Unruh channel. The quantum state of a Dirac field mode qubit in Minkowski spacetime, as observed by
Rob in a constantly accelerated frame, is perceived to be degraded by the Unruh effect.
The Unruh channel for a fermionic  qubit was shown to mimic the effect
of  an  AD channel  (of  rank  2), even  though  the  Unruh effect  is
associated  with  a  non-zero  temperature (suggesting  a  generalized
(squeezed) AD channel, of rank 4).  The damping parameter $\sin r$
attains the maximal value of $\frac{1}{\sqrt{2}}$, and not the maximum
possible value  of 1.   Therefore, under infinite-time  evolution, the
Bloch  sphere  is contracted  to  a finite  volume  rather  than to  a
point. In fact, it tends to an off-centered oblate spheroid that osculates 
with the Bloch sphere at the south pole. This geometric
characterization of the Unruh channel, brings out the points of similarity and
differences between Unruh channels and their conventional counterparts.
Since the  quantum channel capacity is known  to vanish in both
the  fermionic and  bosonic  Unruh effects,  as  does nonlocality,  it
appears that channel capacity  is better reflected in nonlocality than
in entanglement.   We construct the inverse Unruh  channel, namely the
map that returns Rob's noisy qubit  to its initial pure form, and show
that it is NCP. We briefly discuss its status as a physical process.

{\em Acknowledgments.\textemdash}
 SO acknowledges Manipal  University, Manipal, India for accepting
this  work  as  a  part  of  his Ph.D.  thesis.   Authors  SO  and  RS
acknowledge   financial    support   from   the   DST    for   project
SR/S2/LOP-02/2012. We thank the anonymous Referees for their useful comments and
suggestions that helped to improve the paper. 

\bigskip

\vspace{1cm}

\end{document}